\documentclass[10pt,conference,a4paper]{IEEEtran}
\usepackage{times,amsmath,epsfig}
\usepackage{amssymb}
\usepackage{graphicx}
\usepackage{float, array}
\usepackage{hyperref}
\title{A Survey of Cyber Security Countermeasures Using Hardware Performance Counters}
\author{%
{James Christopher Foreman{\small $~^{\#1}$} }%
\vspace{1.6mm}\\
\fontsize{10}{10}\selectfont\itshape
$^{\#}$\,Department of Engineering Fundamentals, University of Louisville \\Louisville, KY. USA
\fontsize{9}{9}\selectfont\ttfamily\upshape
$^{1}$\,jcfore01@louisville.edu\\
\vspace{1.2mm}\\
\fontsize{10}{10}\selectfont\rmfamily\itshape
}
\begin{document}
\maketitle
\begin{abstract} 
Cyber attacks and malware are now more prevalent than ever and the trend is ever upward. There have been several approaches to attack detection including resident software applications at the root or user level, e.g., virus detection, and modifications to the OS, e.g., encryption, application signing, etc. Some approaches have moved to lower level detection and prevention, e.g., Data Execution Prevention. An emerging approach in countermeasure development is the use of hardware performance counters existing in the micro-architecture of modern processors. These are at the lowest level, implemented in processor hardware, and the wealth of data collected by these counters affords some very promising countermeasures with minimal overhead as well as protection from being sabotaged themselves by attackers. Here, we conduct a survey of recent techniques in realizing effective countermeasures for cyber attack detection from these hardware performance counters.
\end{abstract}

\section{Introduction}\label{Introduction}
Cyber security has been at the forefront of mainstream media for several years now as a critical problem for our society to overcome. Attackers are increasingly motivated and enabled to compromise software and computing infrastructure. Cyber security countermeasures are of prime interest in mitigating such attacks and associated malware.

There are many types of countermeasures that are built as software applications, e.g., virus checkers, based on controlling physical access, e.g., biometrics, or enforced as policies, etc. Our investigation is to survey the state of the art in the utilization of Hardware Performance Counters (HPC) to build cyber security countermeasures. HPCs are a promising new resource to address the limitations of typical software, and other countermeasures.

Hardware performance counters are special purpose registers and logic incorporated in the micro-architecture of modern processors and CPUs. They are typically used as debugging tools that run at the lowest level, i.e., on chip, for performance tuning and analysis by collecting information on processor events and the running processes. As the name implies, HPCs are used to count events, such as cache misses, and aid in timing events, such as counting CPU cycles per unit time. This information that is typically used to debug software can now also be used to detect cyber attacks. Their residence in micro-architecture, i.e., in silicon, is a safeguard against their tampering.

\subsection{Recent Related Surveys}
Several related surveys have been performed, e.g., \cite{demme2013feasibility} examines the feasibility of using HPCs to detect malware with several specific examples of HPC data triggers and detection techniques, and others that focus on Control Flow Integrity\cite{de2017survey} (CFI), hardware trojans\cite{bhunia2014hardware}, and side-channel timing attacks\cite{ge2016survey}.  Our survey updates the current state of knowledge and focuses on HPCs in particular, examining several examples and categorizing them by method and attack vector.

\subsection{Using Hardware Performance Counters as Countermeasures}
Hardware performance counters afford a highly granular and low footprint method of detecting anomalous behavior. HPCs reside on the processor chip, implemented in dedicated hardware, so they typically consume minimal resources from the processor. Their inclusion by major processor vendors alleviates the need to develop custom IP cores for cyber attack detection. HPCs collect a wealth of information such as cache misses, event timing, branch mis/predictions, etc. about the running processes. They also execute at the kernel/hardware privilege level, and are difficult to spoof or sabotage by attackers due to their physical persistence in the micro-architecture.

Table \#1 lists some of the commonly used HPCs. Many additional HPCs are available depending on the processor manufacturer, e.g., Intel\cite{intelsystemguide2016}. This table is more thoroughly discussed with supporting data collected from anomaly testing in \cite{garcia2015anomaly}.

\begin{table}[h!]
\caption[Typical hardware performance counters]{Typical hardware performance counters, \emph{adapted from \cite{garcia2015anomaly}}.}
\begin{center}
\begin{tabular}{| l || l || l |}
\hline
cpu-cycles & L1-dcache-loads & dTLB-loads \\ \hline
branches & L1-dcache-stores & iTLB-loads \\ \hline
instructions & L1-icache-loads & dTLB-load-misses \\ \hline
branch-misses & L1-icache-load-misses & iTLB-load-misses \\ \hline
branch-loads & LLC-loads & dTLB-stores \\ \hline
branch-load-misses & LLC-load-misses & dTLB-store-misses \\ \hline
cache-references & LLC-stores & \\ \hline
cache-misses & LLC-store-misses & \\ \hline
ref-cycles & & \\ \hline
bus-cycles & & \\ \hline
\end{tabular}
\end{center}
\label{HPC-table}
\end{table}%

The \emph{perf} utility in Linux is an example method of access. Direct access through machine coding, e.g., inline in C, and custom monitoring software are possible as well. Software development tools should allow HPCs to be activated without source code modification or in some cases rebuilding. In this case, HPCs are in contrast to code instrumentation as they exist to passively and externally monitor the processor behavior. The wealth of data from HPCs lends itself to the discovery of anomalous behavior that is an indicator of a potential attack. HPCs employ one or more of the approaches in detecting attacks. 

\begin{enumerate}
\item Signature based: HPCs collect information about the suspect process and determine if this information corresponds to either known attacks, e.g., \emph{blacklist}, or known safe applications, e.g., \emph{whitelist}. This is similar to approaches used by many virus scanning applications. The whitelist, if practical, has the added benefit of denying any activity that has not been validated, thus mitigating unknown and zero day attacks.
\item Heuristic based: HPCs monitor the suspect process to determine if behavior is anomalous, such as if there are a high number of cache misses or a high number of branch mis-predictions (above a heuristic threshold) to indicate a potential attack.
\item Advanced approaches: HPC data are analyzed and used in more advanced statistical analysis, machine learning, or other artificial intelligence approaches with supervised or unsupervised learning.
\item Hybrid approaches: A combination of one or more of these, possibly also in cooperation with other security countermeasures.
\item Context sensitivity: In addition to monitoring blacklist, whitelist, and heuristic behavior, the context in which the application is running can be part of the classification. This can be realized when the countermeasure creates a Control Flow Graph (CFG) during initial configuration and then monitors when syntactically-correct, though functionally invalid, paths are attempted, such as during code reuse attacks.

\end{enumerate}

The selection of an approach depends on the application and environment. Forming signatures requires specific knowledge of the attack to form a blacklist, or knowledge of all valid (acceptable) applications to form a whitelist. Heuristics are used when this knowledge is less specific, and general knowledge of trends are available through monitoring of the system to set guidelines, e.g., thresholds. Machine learning becomes a better alternative when the system needs to adapt to unknown threats or the execution environment is too dynamic to predict anomalous behavior.

\subsection{Notes for IoT and Embedded Systems}
Embedded systems and systems that comprise the Internet of Things (IoT) usually have the characteristics of limited resources, such as memory, processing power, and network bandwidth. IoT specifically may also include high deployment where many devices are managed. The use of HPCs for countermeasures are still a viable alternative for these, perhaps more so due to the low overhead of HPCs, though the following points should be considered.

\begin{enumerate}
\item Some embedded systems may have limited HPCs available, especially in custom or application specific implementations.
\item The use of black/white lists may require too much storage and the use of machine learning algorithms may require too much processing power. Heuristic approaches tend to work best, though when used alone they may not provide adequate protection.
\item In deployments with many devices, a centralized database or machine learning engine may be able to offset some of the local limitations to provide good protection, providing that network bandwidth is available. Distributed approaches may alleviate limitations when a centralized authority is not practical.
\item Many embedded systems only run a limited selection of applications and/or have static configurations. A whitelist may be more practical and effective in these cases.

\end{enumerate}

\subsection{Notes for Cloud Usage}
Cloud usage and usage in Virtual Machines (VM) should be possible as most VM hypervisors have the option of enabling virtual HPCs. Cloud providers would need to enable this functionality as it is usually not enabled by default. Otherwise, the use of HPCs for countermeasures should be largely transparent to the cloud provider and users. When HPCs are enabled in VMs, it should be ensured that the HPC values presented to the VM OS are only for that VM's activities, which is usually the case and again, managed by the hypervisor. Cloud providers may choose to enable these methods rather than rely on users' requests.

\subsection{Structure of this Paper}
Section \ref{Introduction} introduces the topic of cyber attack detection via HPCs, discusses similar surveys, and includes notes on specific application areas. Section \ref{attacktypes} discusses the types of attacks, i.e. attack categories, including their capabilities and how they are carried out. Section \ref{examples} analyzes several example cyber attack countermeasures using HPCs, categorizing these by countermeasure approach. Section \ref{comparison} provides a summary of this analysis with insights into countermeasure characteristics, implementation, and hybridization of multiple countermeasures that may be utilized for more complete detection coverage while mitigating false positives and false negatives. Finally, Section \ref{future} discusses future directions for HPC-based countermeasures.

\section{Types of Cyber Attacks and Attack Vectors}\label{attacktypes}
\begin{figure*}[h!]
\begin{centering}
\includegraphics[width=\textwidth]{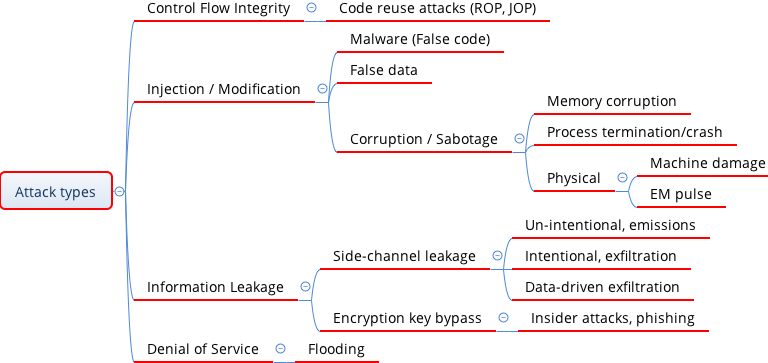}
\caption{Types of cyber attack.}
\label{attack-types}
\end{centering}
\end{figure*}

The types of cyber attacks possible have been well covered in the literature. A brief summary of cyber attack categories is provided in Fig.\ \#\ref{attack-types}.

Code Reuse Attacks (CRA) that compromise control flow integrity seek to alter the normal control flow of a software application to perform malicious activities. Examples include Return Oriented Programming (ROP), in which the attacker gains control of the call stack to rewrite the return address from a function call, and Jump Oriented Programming (JOP), in which the attacker maliciously uses the jump instruction to piece together malicious code fragments. The attack uses existing instructions in executable code or resident libraries that are chained together to form gadgets. These gadgets are similar to functions, i.e., sets of instructions, that are used to perform the malicious activity of the attacker. In most cases, the attacker needs to know the executable code and libraries from which to select gadgets. For commodity operating systems and applications, these are known to the attacker. Address space layout randomization, e.g., code randomization, is an effort to make this more difficult. Also, side channel leaks may allow an attacker to uncover enough information from which to build useful gadgets anyway. Such CRA mitigation has been a primary focus of HPC-based countermeasures as the HPC information collected, such as cache misses, branch mis-predictions, etc., are good heuristic indicators of CRA where control flow becomes detectably anomalous.

False Code Injection (FCI) and modification attacks seek to inject a malicious software payload or overwrite existing application code with such a payload to perform malicious activities. Many of these are done via buffer overflows, and may be performed by other various means. In some cases, false data may be injected to alter program behavior, such as false sensor readings in process control systems. The goals of such attacks may be to seize control, sabotage, or to damage the system being attacked so as to interfere with the performance of its mission. HPC countermeasures for these generally look for anomalous behavior, i.e., contrary to the valid functioning of the application software. Depending on the code overwritten, the counts for various errors may dramatically increase in a short time, e.g., buffer overflow events. 

Information leakage attacks seek to steal information from the target system. Usually, these are passwords or other secrets, and may be executable code fragments in preparation for a code reuse attack. Side channel leakage is the most common vector using cache operation attacks such as flush+reload\cite{yarom2014flush+}, evict+time, prime+probe, and evict+reload. HPCs can detect these from excessive cache misses. HPCs may also monitor event counts for correlation with secret keys when attackers seek to employ HPCs in side channel attacks. Due to the high level of detail HPCs can provide, some attackers may exploit HPCs, for example, to leak the secret key when encryption operations are performed. Martin et al. \cite{martin2012timewarp} have proposed disabling or adding noise to HPCs to reduce their accuracy and subsequent efficacy in an effort to prevent attackers from leveraging these.

In other scenarios, more specific hardware events such as memory corruption by \emph{rowhammer}, which repeatedly accesses (hammers) RAM in a very atypical manner to induce errors in adjacent memory cells, may occur and be detected by HPCs acting as hardware monitors based on RAM access. Some Denial of Service (DoS) attacks may also be detected by HPCs noting that most event counts for normal operation often differ greatly from operation during a DoS, which is typically characterized by extremely high activity.

\section{Approaches to Attack Detection}\label{examples}
From the list of approaches for attack detection in the previous section, several specific examples are examined to establish the current state of the art in HPC cyber security countermeasures. 

\subsection{Signature Based Examples}\label{sig-counter}
Three signature based examples, SIGDROP\cite{wang2016sigdrop}, ConFirm\cite{wang2015confirm,wang2016malicious}, and another by Chiappetta et al.\cite{chiappetta2016real} are examined in their use of HPCs to detect cyber attacks. SIGDROP focuses on detecting Return Oriented Programming (ROP) attacks using two characteristics of such attacks. The first is a high level of mis-prediction by the Return Address Stack (RAS) due to the attackers mis-direction in returns. The second characteristic is that of calls to functions that are very short in instruction length, i.e., gadgets, that are artificially crafted from existing code to perform attack functions. Many such gadgets must be chained together to perform useful work for the attack, thus long chains of very short functions are another signature. Recent studies show that most ROP gadgets have fewer than 6 instructions \cite{pappas2013transparent, cheng2014ropecker, kayaalp2013scrap} and may require chaining of dozens to hundreds of gadgets to perform an attack function. SIGDROP configures hardware performance counters to count if the number of consecutive return address predictor misses is above a threshold, $T_M$, and compares this to the total number of return instructions, $N_R$. If these are nearly equal, then the return address predictor is missing almost all the time, which is one of the characteristics of a ROP attack. A HPC is also configured to count total instructions executed, $N_I$, to check the average number of instructions per missed return address prediction. Noting that the typical number of instructions per gadget or per return is $T_I \le 6$ for ROP attacks, the second signature is found by $N_I \le (T_I \times T_M)$, which is true when the total number of instructions is less than or equal to the typical ROP gadget length times the number of return address predictor misses. Thus, SIGDROP is an example of a blacklist signature approach. The blacklist behavior is determined by comparison against known attacks.

ConFirm\cite{wang2015confirm,wang2016malicious} uses hardware performance counters to detect malicious software either injected into firmware or by performing CRAs using firmware code. ConFirm is a whitelist signature approach since firmware is known in advance and rarely changes. ConFirm performs HPC checks at various points in the firmware code execution process to determine if configured HPCs are at typical values. Since the same code always executes under normal circumstances, these should be very consistent. An attack would introduce new operations and thus change the HPCs. The whitelist behavior is determined by profiling the valid code (firmware) prior to deployment (offline) to collect good HPC values and determine optimal checkpoints.

Chiappetta et al.\cite{chiappetta2016real} proposed using HPCs to detect side channel attacks, specifically on cache memory to compromise encryption through information leakage, such as flush+reload. The countermeasure employs a utility, \emph{quickhpc}, that allows the HPC to be queried much faster, at microsecond resolution. Under normal circumstances, the encryption process would be expected to benefit from the cache for a significant portion of the process time. However, when under attack the encryption process never benefits from the cache, because the flush+reload side channel attack is constantly flushing the cache and timing the reload to determine program flow of the encryption process. These cache misses are collected by the HPC which can signal anomalous operation. If using simple threshold heuristics, there could be many false positives, so Chiappetta employs simple machine learning to determine a signature for the encryption process. This is another example of a whitelist approach that uses unsupervised training.

\subsection{Heuristic Based Examples}\label{heu-counter}
Heuristic based examples, such as ANVIL\cite{aweke2016anvil}, CacheShield\cite{briongos2017cacheshield}, by Lui et al.\cite{liu2014leveraging}, by Torres et al.\cite{torres2016can}, and Eunomia\cite{yuan2011security}, provide direct detection of attacks when certain events count past preset thresholds, either individually or in some combination. These tend to perform better when the effects of attacks are more generally known, e.g., when jump oriented attacks result in high branch mis-predictions. These may result in a higher number of false positives depending on the process being executed, especially when there is a wide range of potentially valid processes. However, they are simple to implement and can act as a pre-filter for more advanced and resource consuming detection approaches. ANVIL is a Linux-based kernel module to mitigate rowhammer attacks, specifically new forms of rowhammer that seek to evade simple rowhammer countermeasures that DRAM manufacturers are now employing, such as on-DRAM caches. ANVIL works by monitoring the locality of DRAM row accesses out of the LLC misses (\textsc{longest lat cache.miss}). Once a preset threshold of LLC misses is exceeded, a second stage of detection samples virtual addresses for a time duration using Load Latency (\textsc{mem trans retired.load latency}) and Precise Store (\textsc{mem trans retired.precise store}) events to determine locality. Once an attack is detected, the rows adjacent to the rows being attacked are refreshed through a read operation. This is only performed as needed so that false positives have very little effect on the system. Thresholds can be determined by observation of bit flips, and may also be empirically set based on DRAM specifications.

CacheShield\cite{briongos2017cacheshield} is designed to be a user-level tool, with low performance impact for legacy systems, that specifically targets cache attacks. Cache misses, a common symptom of cache attacks, are counted using various cache miss HPCs. CacheShield is configured by monitoring known valid and malicious applications to determine cache miss thresholds for detection, and selects the specific HPCs that are most affected for the application. The example given in the paper was for OpenSSL and the L3 cache. A cache attack is detected when an abrupt change in the statistical distribution of cache misses occurs.

Lui et al.\cite{liu2014leveraging} developed a countermeasure to stack buffer overflow attacks used to compromise control flow integrity. A two-level approach is used with the first level being a heuristic pre-filter to facilitate low overhead on embedded systems. Stack buffer overflow attacks redirect control flow through dynamically overwriting the return address of a procedure, which results in instruction cache misses and mis-prediction of return addresses. Anomalous behavior is detected when these occur above an established threshold.

Torres et al.\cite{torres2016can} investigated if data-only exploits could be detected at runtime with HPCs. Examples of data oriented attacks are SQL injections or any other mis-information whereby malformed data sent to a host causes the host to disclose secret information. The Heartbleed attack, studied specifically in this work, uses an overestimate of the size of keep-alive packets that keep secure channels open, causing the host to respond with extra data, which contain sensitive information.

Eunomia\cite{yuan2011security} is another example of earlier work that is similar to these where deviations in PMU-event counts signal malicious activity versus valid processes. This paper includes a good quantitive discussion of HPC deviation values in general under attack scenarios for reference. 

\subsection{Examples of Machine Learning and Context Sensitivity}\label{mach-counter}
Machine learning includes most approaches in the area of artificial intelligence. Learning may be supervised, such as training HPC data against known valid and known malicious applications. This learning is usually offline, i.e., the classification engine for detecting malicious behavior is developed before runtime or deployment. Learning may also be unsupervised, such as online during runtime based on accumulated information, e.g., information from HPCs. Security policy may still be specified for unsupervised learning and the classification engine will learn violations to this policy. Context sensitivity implies knowledge of the operating environment or application. This knowledge may include information from the source or binary code such as the proper execution paths, e.g., control flow graph verification, or mathematical rules, such as those extracted from the code or based in physics for physical processes, to validate proper operation of the compiled application. Instrumentation of the binary may be performed to provide checkpoints within the application to facilitate these checks. Knowledge of the user environment may be used to detect deviation from expected user behaviors, or even the behaviors of the machine hardware. 

The goal of machine learning is to provide a more advanced detection scheme that eliminates the false positives from simple signatures and heuristics as well as eliminating the false negatives when sophisticated attacks are launched that use valid code fragments and other seemingly valid approaches. Machine learning is typically of much higher processing overheads and is often deployed as a second layer to a signature or heuristic first layer, which acts as a pre-filter to minimize the performance impact.

Torres et al.\cite{torres2016can} performed a survey of approaches that were essentially intelligent outlier rejection. The desired approach characteristic was unsupervised learning by using collected HPC data only, i.e., a data-driven approach. Cache misses and branch mis-predictions were common variables studied. During runtime, HPC data was collected for specific intervals (1ms, 10ms, 100ms) with the assumption that valid activity was more common (normal) and that invalid activity (attacks) would be statistical outliers to the HPC data. The countermeasure behaved similarly to heuristic analysis without the necessity of pre-determining heuristic thresholds. The machine learning portion would build a model in memory of the valid state space as the statistical norm.

HPCMalHunter\cite{bahador2014hpcmalhunter} dynamically monitors HPC data to classify malicious behavior. This approach uses supervised learning and offline pre-training to build a database for classification. The HPC data assembled into vectors (monitored) in the example were: Branch instructions retired (BIR), load instructions retired (LIR), store instructions retired (SIR), and mis-predicted branch instructions (MBI). The database is a matrix and HPC event data is formatted as a vector input for classification, similar to an artificial neural network except by a Support Vector Machine\cite{wiki2018svm} (SVM) in this case. As HPC data are typically very sparse, the SVM matrix is optimized by Single Value Decomposition (SVD) to reduce its dimensionality and reduce storage and processing overheads resulting in the final classification engine. This particular countermeasure examined HPC data in blocks of 100,000 machine instructions, and the span of examining multiple HPC data (BIR, LIR, SIR, MBI) facilitated a very low false positive rate.

Nomani et al.\cite{nomani2015predicting} developed an Artificial Neural Network (ANN) classifier to determine the \emph{phase} of a running application as a countermeasure against side channel attacks. Here, side channel attacks refer to attacks that attempt to capture secret information or perform malicious activities on shared resources. Phase refers to the types of resources and functional units that are utilized at that time, such as a memory phase during high memory accesses, a floating point phase during floating point operations, an integer phase, etc. When multiple applications, or an application and a malicious program, share resources, the potential for an attack is much higher\cite{zhang2012cross}, and thus monitoring should be more vigilant. The ANN provides a black-box approach to determining the phase in which running applications reside or are in transition and purposely influences the OS scheduler to avoid scheduling other applications on the same processor using the same functional resources. Contrary to increasing overhead, the countermeasure on average reduced resource load by as much as 25\% in some cases as a side benefit. Once trained via supervised learning, the ANN was able to perform classification well under the average time between context switches allowing the scheduler sufficient time to recalculate thread scheduling in most cases.

Alam et al.\cite{alamperformance} developed a countermeasure that employed two novel methods. The first was consideration that lots of HPC data were known or could be generated for valid applications, while HPC data for attacks were rare or would be unknown due to zero-day attacks. Therefore, a single-class SVM was developed to only classify valid behavior. The failure to classify valid behavior determined potential invalid (malicious) behavior. The behavior was further analyzed to select the most likely HPC variables for attack classification based on how the anomalous behavior deviated from valid behavior. The second novel method used in the countermeasure was Dynamic Time Warping\cite{wiki2018dtw} (DTW). HPC data represent time series of various event counts, such as cache misses. A side channel attack may seek to exploit HPCs by superimposing the secret key or other sensitive information on these time series through seemingly benign operations to exfiltrate the sensitive information. Therefore, these time series (HPC events) are monitored and correlated with sensitive information to see if there is a match. The DTW algorithm allows detection even when the time series is compressed, stretched, or scaled with respect to the sensitive information pattern.

BehavioR based Adaptive Intrusion detection in Networks\cite{jyothi2016brain} (BRAIN) is a countermeasure for distributed Denial of Service (DoS) attacks in networks. Most network-based DoS countermeasures use heuristics on network traffic by examining packets for attack signatures or specific attack behavior. BRAIN enhances this by adding HPC data in the analysis of DoS attacks under the assumption that processors also behave differently during such attacks. BRAIN is trained during idle and normal operation as well as during known DoS attacks, i.e., supervised and online. Network heuristics from traditional approaches are combined with BRAIN's HPC-based information via unsupervised K-means clustering that is then used to form a SVM for final classification. Claimed results are zero false positives with 99.8\% true positive detection, conditional on the span of the DoS attack scenarios used in training. 

FlowGuard\cite{liu2017transparent} is a countermeasure approach worth mentioning here although it does not use HPCs. It does, however, utilize Intel Processor Trace\cite{intel2013processortrace}, a debugging tool also implemented in micro-architecture. FlowGuard uses machine learning of control flow paths to form a valid Control Flow Graph (CFG). It then compresses the CFG information in the same format as that supplied by Intel Processor Trace to allow rapid, direct comparison of runtime control flow with these learned valid paths. Paths are ranked with the most common paths ranked highest. During runtime, deviation from valid paths will indicate an anomaly and potential attack that can then be examined with additional analysis, such as a hybrid approach with HPCs.

\section{Comparison of Countermeasure Approaches}\label{comparison}
\begin{figure*}[h!]
\begin{centering}
\includegraphics[width=5in]{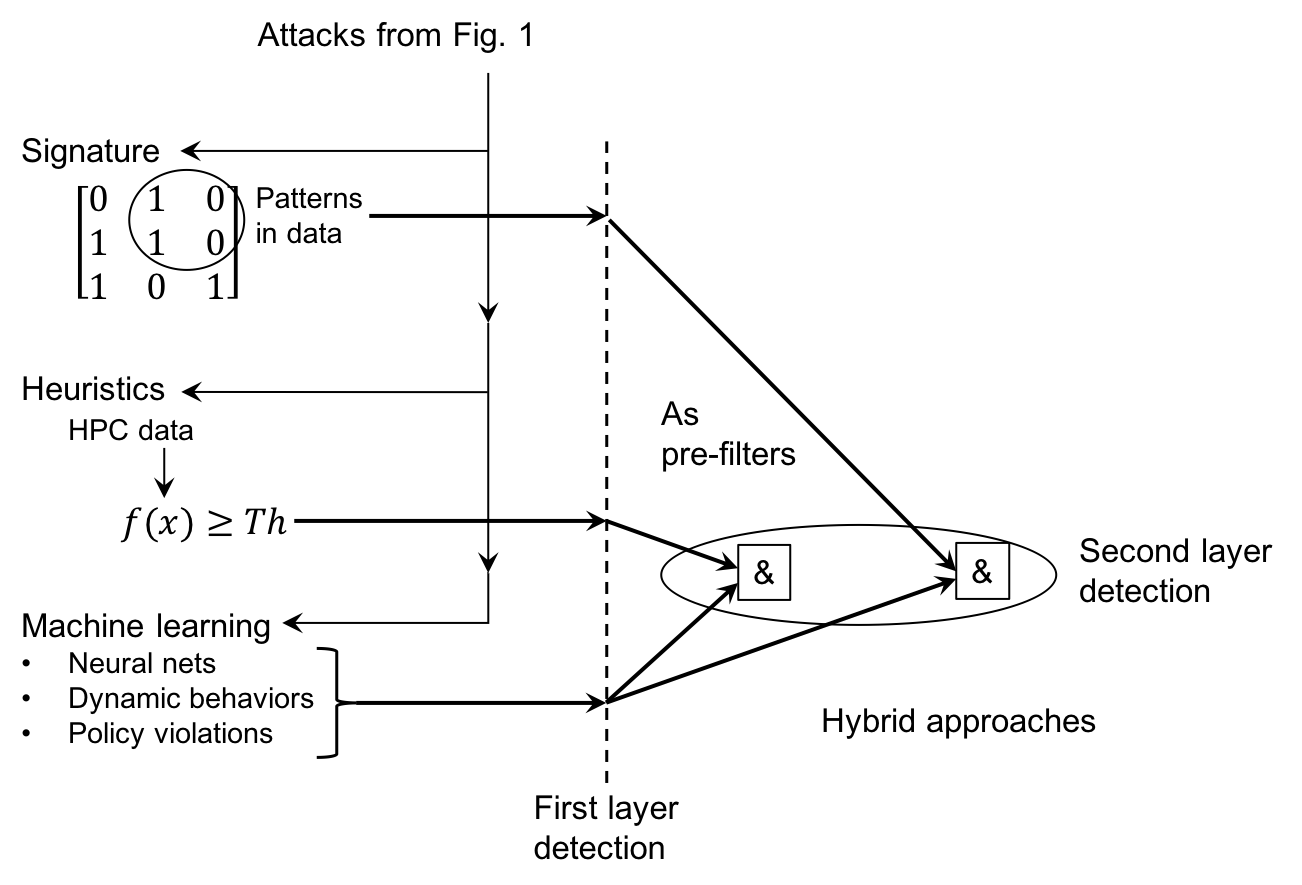}
\caption{Summary of HPC countermeasures.}
\label{summary-fig}
\end{centering}
\end{figure*}

In this section, a comparison of countermeasures approaches as exemplified in Section \ref{examples} is given. Table \ref{exampletabulation} tabulates the examples given in Section \ref{examples} with respect to name and citation, HPCs utilized, the general category also from Section \ref{examples}, and the types of attacks for which that example is good for detecting. Table \ref{categorycomparison} provides a comparison of the general categories with respect to characteristics of countermeasures within that category and application notes that fit that category. Figure \ref{summary-fig} illustrates the process flow of countermeasure categories and how multiple approaches may be used in hybrid configurations.

\begin{table*}[h!]
\fontsize{10pt}{12pt}\selectfont
\caption{Examples of HPC countermeasures.}
\begin{center}
\begin{tabular}{|>{\raggedright}p{0.15\linewidth}|>{\raggedright}p{0.15\linewidth}|>{\raggedright}p{0.3\linewidth}|p{0.3\linewidth}|}
\hline
\textbf{Name} & \textbf{Category} & \textbf{HPCs used} & \textbf{Attacks targeted} \\ \hline \hline

SIGDROP \cite{wang2016sigdrop} & signature & return address predictor misses, total number of return instructions & ROP and other CRAs \\ \hline
ConFirm \cite{wang2015confirm,wang2016malicious} & signature & various HPCs, code instrumentation via checkpoints & malicious software injected into firmware or CRAs using firmware \\ \hline
Chiappetta et al. \cite{chiappetta2016real} & signature & cache misses and high speed HPC query & side channel attacks, cache memory, information leakage, e.g., flush+reload \\ \hline \hline

ANVIL \cite{aweke2016anvil} & heuristic & LLC misses, Load Latency, Precise Store & rowhammer \\ \hline
CacheShield \cite{briongos2017cacheshield} & heuristic & various cache misses & cache attacks on legacy systems \\ \hline
Lui et al. \cite{liu2014leveraging} & heuristic & instruction cache misses and mis-prediction & stack buffer overflow attacks \\ \hline
Torres et al. \cite{torres2016can} \textdagger & heuristic & Cache misses and branch mis-predictions & general, anything outside the norm, Heartbleed \\ \hline
Eunomia \cite{yuan2011security} & heuristic & PMU events & general \\ \hline \hline

Torres et al. \cite{torres2016can} \textdagger & ML, OR & Cache misses and branch mis-predictions & general, anything outside the norm, Heartbleed \\ \hline
HPCMal-Hunter\cite{bahador2014hpcmalhunter} & ML, SVM, SVD & BIR, LIR, SIR, MBI & general \\ \hline
Nomani et al. \cite{nomani2015predicting} & ML, ANN & HPCs by resource (memory, floating point, integer, etc.) & general \\ \hline
Alam et al. \cite{alamperformance} & ML, SVM, DTW & chooses HPCs from learning & general \\ \hline
BRAIN \cite{jyothi2016brain} & ML, k-means, SVM & chooses HPCs from learning & distributed denial of service \\ \hline 
FlowGuard \cite{liu2017transparent} & ML, CFG & Intel Processor Trace & ROP and others that alter process flow \\ \hline \hline

\end{tabular}
\end{center}
\label{exampletabulation}
\end{table*}%

The following terms are used in Table \ref{exampletabulation}. Branch instructions retired (BIR), load instructions retired (LIR), store instructions retired (SIR), and mis-predicted branch instructions (MBI), Processor Management Unit events (PMU), Code Reuse Attacks (CRA), Machine Learning (ML), self-directed Outlier Rejection (OR), Support Vector Machine (SVM), Single value Decomposition (SVD), Artificial Neural Network (ANN), Dynamic Time Warping (DTW), Control Flow Graph (CFG), and Return Oriented Programming (ROP). General attack effectiveness, usually in machine learning, implies the countermeasure is used to detect general attack behavior versus a specific class. Chooses by learning implies the countermeasure selects HPCs that are best suited, i.e. most affected, by the attack class to be detected.
\\ \textdagger Torres et al. incorporates both heuristic and machine learning aspects to its approach. 

\begin{table*}[h!]
\fontsize{10pt}{12pt}\selectfont
\caption{Comparison of HPC countermeasure categories.}
\begin{center}
\begin{tabular}{|p{0.12\linewidth}|p{0.45\linewidth}|p{0.35\linewidth}|}
\hline
\textbf{Category} & \textbf{Characteristics/requirements} & \textbf{Application (Use cases)} \\ \hline \hline

Signature & memory intensive, not adaptive, training of signatures, simple & specific/known attacks and apps, blacklist/whitelist \\ \hline
Heuristic & minimal footprint, generally quickest, tuning of thresholds, false posi/negatives may be higher & when general behaviors are known, attack exceeds some threshold \\ \hline
Machine learning & generally largest footprint, potentially best with minimal false posi/negatives, context sensitivity & dynamic/unknown attack and app behaviors, deep attacks, zero day attacks  \\ \hline
Hybrid & large footprint offset by filtering (multi-layer), minimal impact with minimal false posi/negatives & when the combined features of two or more countermeasure layers is beneficial \\ \hline

\end{tabular}
\end{center}
\label{categorycomparison}
\end{table*}%

In Fig.\ \ref{summary-fig}, attacks of all types, as denoted in Fig.\ \ref{attack-types}, enter the target system. Signature countermeasures, Section \ref{sig-counter} scan attack activity in various HPCs for specific patterns. The assumption is that known attacks impact various counters in a predictable and repeatable manner. Another approach may be the use of heuristics, Section \ref{heu-counter}. Heuristics look for anomalous activity in HPCs as the exceeding a preset threshold. These are usually quick and simple in implementation. Machine learning approaches, Section \ref{mach-counter}, may be utilized through any number of more advanced approaches. Any one of these may be used as a first layer of detection of cyber attack. Hybrid approaches will use one or more of these as a pre-filter in combination with one or more of these at a second layer of detection in an effort to mitigate false positives/negatives, incorporate intelligent approaches, and minimize resource requirements by reserving complex countermeasure activities only after passing through simpler pre-filtering. A typical hybrid approach would be using a heuristic as a pre-filter at the first layer and (\&) using machine learning for further analysis at the second layer.

\section{Future Directions}\label{future}
From this survey, hardware performance counters have already been used in a variety of cyber attack countermeasure approaches. In most cases, they are an existing resource and implemented in on-chip in separate hardware, thus minimizing their impact both on application performance and in application development.

The motivations for including HPCs in micro-architecture have been towards improved debugging and application stability. If additional micro-architectural features are incorporated that have cyber security as a primary mission, more approaches may be possible in the near future. Additional micro-architectural approaches already available include execution and information flow monitoring, e.g., Intel Processor Trace\cite{intel2013processortrace}, built in self tests, subroutines from Joint Test Action Group (JTAG) interfaces, and others. Also of note are the increasing reliance on cloud systems and computing as a service. In the same method of using micro-architectural approaches, hypervisor based approaches should also be examined. Research should continue in such approaches as an additional front in the prevention of cyber attacks, especially noting the ever increasing footprint of software based approaches. As the number and complexity of attacks increases, these applications utilize more resources and become more difficult to develop and manage.

While cyber attacks are increasing in number and sophistication, the vast majority of these are still not able to perform much malicious activity without leaving basic hardware signatures, such as missed branch predictions, cache misses, hammering of rows, etc. HPC based countermeasures need to move out of research and into to the mainstream of attack detection software as quickly as possible. In cases where a hardware/software system configuration is completely known, HPCs could theoretically detect most any direct attack. In many cases, side channels can be detected as well, e.g., Dynamic Time Warping.

\subsection{Special Note Meltdown and Spectre}
Currently, the cyber attacks of Meltdown and Spectre \cite{cert2018meltdownspectre} have emerged at the micro-architectural level. In Meltdown, an attack attempts an unauthorized read of privileged memory, to which it is not allowed access. Though the processor will eventually deny access to this memory, it will still fetch and in most cases perform some processing, i.e., speculative processing, with this memory. The attacker then attempts to intercept this information or the result prior to failing the privilege check, or in some cases as a residual after the privilege check. HPCs that count privilege check violations, if developed, might indicate this attack. Spectre is a more generalized class of vulnerabilities similar to Meltdown, focusing on branch prediction. In speculative execution schemes, both branch options may be followed until the correct branch is finally determined. The processor would then discard the mis-predicted branch, though side effects of this would remain and encourage a side channel attack. These types of branch mis-predictions are common whether the attack is present or not, i.e., the attack does not cause these mis-predictions. Some mitigations involve preventing out of order execution for vulnerable processes, but this carries significant performance impacts. Perhaps future work in HPCs as cyber attack countermeasures could employ multiple event counts to selectively classify Meltdown and Spectre attacks since HPCs work at the micro-architectural level with these attacks.

\bibliographystyle{IEEEtran}
\bibliography{HPC-Survey-IJCSIS}

\end{document}